\documentclass[aps,prl,reprint,showpacs]{revtex4-1}
\usepackage{graphicx} 
\usepackage{tabularx}
\usepackage{dcolumn} 
\usepackage{color}
\usepackage{bm}
\usepackage{amsmath}
\usepackage{amssymb}
\newcommand{\bi}[1]{\ensuremath{\boldsymbol{#1}}} 
\newcommand{\nn}{\nonumber}
 
\begin{document} 
 
\title{ 
Edge mass current and roles of Majorana Fermions in superfluid $^3$He A-phase
}

\author{Yasumasa Tsutsumi} 
\affiliation{Department of Physics, Okayama University, 
Okayama 700-8530, Japan} 
\author{Kazushige Machida} 
\affiliation{Department of Physics, Okayama University, 
Okayama 700-8530, Japan} 
\date{\today}

\begin{abstract} 
The total angular momentum associated with the
edge mass current flowing at the boundary in the superfluid $^3$He A-phase confined in a disk
is proved to be $L=N\hbar/2$, consisting of $L^{\rm MJ}=N\hbar$ from 
the Majorana quasi-particles (QPs) and $L^{\rm cont}=-N\hbar/2$ from the continuum state.
We show it based on an analytic solution of the chiral order parameter for quasi-classical Eilenberger equation.
Important analytic expressions are obtained for mass current, angular momentum, and density of states (DOS).
Notably the DOS of the Majorana QPs is exactly $N_0/2$ ($N_0$: normal state DOS)
responsible for the factor 2 difference between $L^{\rm MJ}$ and  $L^{\rm cont}$.
The current decreases as $E^{-3}$ against the energy $E$, and $L(T) \propto -T^2$.
This analytic solution is fully backed up by numerically solving the Eilenberger equation.
We touch on the so-called intrinsic angular momentum problem.
\end{abstract} 
 
 
 
\maketitle 


The superfluid  is a paradigmatic and fertile testing ground to
investigate the central questions concerning the
origin of anisotropic superfluidity since the $p$-wave pairing symmetry 
for the  $^3$He A-phase and B-phase is firmly established~\cite{vollhardt:book}.
Recently the topological aspect of the $^3$He A- and B-phases 
has become a focus because there has been no other concrete
topological superfluids or superconductors established so far.
The topological superfluid defined by a topological number~\cite{hasan:2010}.
At an edge of the superfluid $^3$He faced to a topologically trivial vacuum,
a superfluid gap is closed by a topological phase transition at the interface.
The surface Andreev bound state of Majorana Fermions must appear~\cite{read:2000}.
The topological features are quite different in the $^3$He A- and B-phases.
The $^3$He A-phase is a chiral superfluid with the spontaneous edge mass current
while B-phase is a helical superfluid with the edge spin current~\cite{tsutsumi:2011b}.
The Majorana nature for the A- and B-phases is also distinctive;
Majorana Fermions have the surface Andreev bound state with linear 
dispersion relation centered at  the zero energy 
forming a ``Majorana valley'' in the A-phase~\cite{tsutsumi:2010b,tsutsumi:2011b} 
and a ``Majorana cone'' in the B-phase~\cite{schnyder:2008,chung:2009,nagato:2009}.
This Majorana cone is emerging by recent experiments~\cite{murakawa:2009,murakawa:2011}.
However, there has been no firm evidence for detecting the Majorana zero mode.
We are still in the process to better characterize this elusive quasi-particle.

The edge mass current spontaneously generated at
the boundary in the superfluid $^3$He A-phase
is carried both by the Majorana quasi-particles (QPs) 
within the energy gap and by QPs in the continuum state outside the gap.
Those two kinds of QPs play a different role for the edge mass current and
associated total angular momentum $L$ of the whole system, such as a disk.
Stone and Roy~\cite{stone:2004} proved that 
in the A-phase the magnitude of the total angular momentum by the edge mass current is 
$N\hbar/2$ at the zero temperature
($N$ is the total number of $^3$He atoms in disk geometry).

Here we pose problems on what amount of this mass current or
the angular momentum $L$ carried by the Majorana QPs and
the remaining continuum state. 
It will turn out that exactly $L^{\rm MJ}=N\hbar$ for Majorana QPs
and $L^{\rm cont}=-N\hbar/2$ for the continuum state, giving rise to 
$L=N\hbar/2$ for the total angular momentum of the system.
Then it is shown that the excess angular momentum due to the Majorana QPs comes 
from the fact that the density of states (DOS) of the Majorana QPs equals  exactly to $N_0/2$
with the normal state DOS $N_0$ at the Fermi energy $E_F$.
It will be also demonstrated that the contribution of the particles with 
the energy $E$ to the mass current decreases as $E^{-3}$,
namely it is neither confined to the narrow energy shell region around the Fermi level,
nor the whole particles deep inside the Fermi level. It is marginal.

We base our arguments with the quasi-classical Eilenberger theory that is 
valid for $\xi\gg k_F^{-1}$ ($\xi\sim 10$--100 nm  the coherent length and $k_F^{-1}\sim 0.1$ nm)
well satisfied for the superfluid $^3$He.
We first find an analytic solution for the order parameter for the A-phase when 
the system has a boundary. This analytic solution satisfies the Eilenberger equation and gives
useful and transparent information on various physical properties at the boundary.
Since this analytic solution exactly satisfies the Eilenberger equation,
but is not a self-consistent solution for the whole problem unless taking the infinite cutoff energy, we back it up by 
numerically solving this problem self-consistently, finding that 
the analytic solution captures the essential points and the numerical 
solution gives additional insights to further characterize the Majorana QPs.

We will touch on the so-called intrinsic angular momentum (IAM) problem,
a long standing controversy concerning $L$ at the zero temperature.
They are summarized as $L^{\rm IAM}=(N\hbar/2)(\Delta/E_F)^{\gamma }$ 
with $\gamma=0$~\cite{ishikawa:1977,ishikawa:1980,mermin:1980,volovik:1995,kita:1998}, 
$\gamma=1$~\cite{anderson:1961,leggett:1975}, and $\gamma=2$~\cite{volovik:1975a,cross:1975},
where $\Delta$ is a superfluid gap.
Since $\Delta/E_F\sim 10^{-3}$ for the superfluid $^3$He,
the IAM can be observed macroscopically only when $\gamma=0$.
It is our hope that the present investigations may contribute to this IAM problem.




We consider disk geometry with a radius $R$ much larger than the coherence length 
and a small thickness along the $z$-direction to align $l$-vector toward the $z$-direction,
where $d$-vector is known to be also aligned toward the $z$-direction by the dipole interaction~\cite{vollhardt:book}.
This system is realized by confining the superfluid $^3$He A-phase 
in a slab of sub-$\mu$m thickness~\cite{tsutsumi:2011b}
that is already realized experimentally~\cite{bennett:2010}.
Since the $d$-vector is fixed, we consider the orbital part of the order parameter 
$\Delta(x,\bi{k})=A_x(x)k_x+iA_y(x)k_y$ in the one-dimension toward the radial direction $x$,
where $\bi{k}$ normalized by $k_F$ is the relative momentum of a Cooper pair.
The coefficients $A_x$ and $A_y$ can be chosen as a real number without loss of generality.
We assume that the edge at $x=0$ is specular
where only the coefficient $A_x$ is suppressed at the edge and $|A_x|=|A_y|$ far from the edge
because the chiral state is recovered in the bulk.

Microscopic information of the edge mass current is contained by the quasi-classical Green's functions
$g(x,\bi{k},\omega_n)$, $f(x,\bi{k},\omega_n)$, and $\underline{f}(x,\bi{k},\omega_n)$
which satisfy the normalization condition $g^2=1-f\underline{f}$.
The quasi-classical Green's functions are calculated by using Eilenberger equation~\cite{eilenberger:1968} as
\begin{equation}
\begin{split}
\left(\omega_n+\hbar v_Fk_x\frac{\partial }{\partial x}\right)f
&=\Delta(x,\bi{k})g, \\
\left(\omega_n-\hbar v_Fk_x\frac{\partial }{\partial x}\right)\underline{f}
&=\Delta(x,\bi{k})^*g,
\label{eilenberger}
\end{split}
\end{equation}
where $v_F$ is the Fermi velocity and $\omega_n=(2n+1)\pi k_BT$ is the Matsubara frequency with 
$n\in\mathbb{Z}$.
The self-consistent condition for the pair potential is given as the gap equation
\begin{align}
&\Delta(x,\bi{k})=N_0\pi k_BT\nn\\
&\times\sum_{0\le\omega_n\le\omega_c}
\left\langle V(\bi{k},\bi{k}')\left[f(x,\bi{k}',\omega_n)+\underline{f}(x,\bi{k}',\omega_n)^*\right]\right\rangle_{\bi{k}'},
\label{gap}
\end{align}
where $\langle\cdots\rangle_{\bi{k}}$ indicates the Fermi surface average
and $\omega_c$ is a cutoff energy set to be $\omega_c=40\pi k_BT_c$ in later numerics
with the transition temperature $T_c$.
The pairing interaction is given as $V(\bi{k},\bi{k}')=3g_1\bi{k}\cdot\bi{k}'$ for Cooper pairs 
with the orbital angular momentum $l=1$, where $g_1$ is 
a coupling constant for the $p$-wave channel.
We use the relation $(g_1N_0)^{-1}=\ln(T/T_c)+2\pi k_BT\sum_{0\le\omega_n\le\omega_c}\omega_n^{-1}$
and solve Eq.~\eqref{eilenberger} by the Riccati method~\cite{nagato:1993,schopohl:1995} in numerical calculations.

By using the quasi-classical Green's function, the mass current is calculated by
\begin{align}
\bi{j}(x)=mv_FN_0\pi k_BT\sum_{-\omega_c\le\omega_n\le\omega_c}
\langle \bi{k}g(x,\bi{k},\omega_n)\rangle_{\bi{k}},
\label{current}
\end{align}
where $m$ is the mass of the $^3$He atom.
The mass current and local density of states (LDOS) for an energy $E$ are given by
\begin{align}
\bi{j}(x,E)&=\langle \bi{j}(x,\bi{k},E)\rangle_{\bi{k}}\nn\\
&=mv_FN_0\langle\bi{k}{\rm Re}\left[g(x,\bi{k},\omega_n)|_{i\omega_n\rightarrow E+i\eta }\right]\rangle_{\bi{k}},
\label{currentE}\\
N(x,E)&=\langle N(x,\bi{k},E)\rangle_{\bi{k}}\nn\\
&=N_0\langle{\rm Re}\left[g(x,\bi{k},\omega_n)|_{i\omega_n\rightarrow E+i\eta }\right]\rangle_{\bi{k}},
\label{LDOS}
\end{align}
respectively, where $\eta$ is a positive infinitesimally small constant.


In order to better characterize the Majorana QPs,
it is essential to decompose the edge mass current into the Majorana QPs of the Andreev bound state within the energy gap and the continuum state outside the gap.
This is our first task.

\begin{figure}
\begin{center}
\includegraphics[width=5cm]{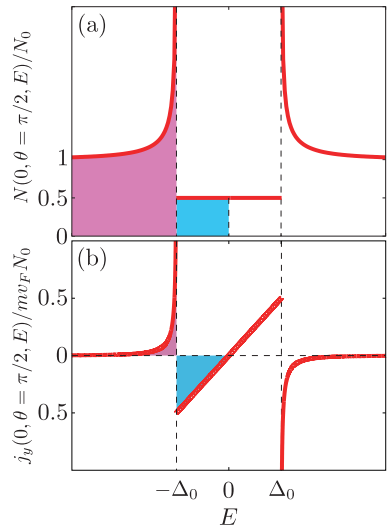}
\end{center}
\caption{\label{energy}(Color online) 
Energy profiles of $\theta$-angle-resolved LDOS (a) and mass current along the edge (b) at $x=0$ for $\theta=\pi/2$.
At the zero temperature, quasi-particles fill the colored (shaded) states in (a).
Mass current from the bound and continuum states is derived 
by integrating the blue (light gray) and pink (gray) regions in (b), respectively.
}
\end{figure}

We solve Eilenberger equation~\eqref{eilenberger} under the pair potential with
\begin{align}
A_x=\Delta_0\tanh\left(\frac{x}{\xi }\right),\ A_y=\Delta_0,
\end{align}
where $\Delta_0$ is the  superfluid energy gap in the bulk and 
the coherent length is defined by $\xi\equiv\hbar v_F/\Delta_0$.
This pair potential is hinted by Ref.~25.
This order parameter form embodies the fact that the $k_x$- ($k_y$-)component 
is suppressed (intact) at the boundary.
The quasi-classical Green's function~\cite{note1}
\begin{align}
&g(x,\bi{k},\omega_n)=\frac{1}{\sqrt{\omega_n^2+\Delta_0^2\sin^2\theta }}\nn\\
&\times\left[\omega_n+\frac{\Delta_0^2\sin^2\theta\cos^2\phi }
{2(\omega_n+i\Delta_0\sin\theta\sin\phi)}\ {\rm sech}^2\left(\frac{x}{\xi}\right)\right],
\label{Green}
\end{align}
is shown to satisfy Eq.~\eqref{eilenberger} after lengthy but simple calculations,
where $k_x=\sin\theta\cos\phi$ and $k_y=\sin\theta\sin\phi$
in the spherical coordinates on the unit Fermi surface.
By the quasi-classical Green's function in Eq.~\eqref{Green}, 
we can calculate $\theta$-angle-resolved LDOS by Eq.~\eqref{LDOS} as
\begin{align}
N(x,\theta,E)\equiv\int\frac{d\phi }{2\pi }N(x,\bi{k},E)
=\frac{N_0}{2}\ {\rm sech}^2\left(\frac{x}{\xi }\right),
\end{align}
for the Majorana bound state $|E|<\Delta_0\sin\theta$ and
\begin{align}
N(x,\theta,E)&=N_0\Biggl[\frac{|E|}{\sqrt{E^2-\Delta_0^2\sin^2\theta }}\nn\\
&-\frac{1}{2}\left(\frac{|E|}{\sqrt{E^2-\Delta_0^2\sin^2\theta }}-1\right)
\ {\rm sech}^2\left(\frac{x}{\xi }\right)\Biggr],
\end{align}
for the continuum state $|E|>\Delta_0\sin\theta$.
Thus, QPs feel the pair potential $\Delta_0\sin\theta$
where $\theta$ is the polar angle.
The $\theta$-angle-resolved LDOS at the edge $x=0$ for $\theta=\pi/2$ is shown in Fig.~\ref{energy}(a).
The LDOS from the Majorana zero energy mode has a constant and finite value $N_0/2$.
Similar value is also obtained by self-consistent calculation in a finite temperature~\cite{tsutsumi:2010b,tsutsumi:2011b}.
This means that the linear dispersion of the Majorana QPs is constant and independent of $\theta$.
According to the numerical calculation~\cite{tsutsumi:2010b,tsutsumi:2011b},
the velocity $v_{\rm MJ}$ of the Majorana QPs is $v_{\rm MJ}\sim\Delta_0/\hbar k_F\ll v_F$.

We can also calculate $\theta$-angle-resolved mass current along the edge by Eq.~\eqref{currentE} as
\begin{align}
j_y(x,\theta,E)\equiv&\int\frac{d\phi }{2\pi }j_y(x,\bi{k},E)\nn\\
=&\frac{mv_FN_0}{2}\frac{E}{\Delta_0}\ {\rm sech}^2\left(\frac{x}{\xi }\right),
\end{align}
for the bound state $|E|<\Delta_0\sin\theta$ and
\begin{align}
j_y(x,\theta,E)=&-\frac{mv_FN_0}{4}\frac{1}{\Delta_0}\frac{E}{|E|}\Biggl[\sqrt{E^2-\Delta_0^2\sin^2\theta }\nn\\
&+\frac{E^2}{\sqrt{E^2-\Delta_0^2\sin^2\theta }}-2|E|\Biggr]\ {\rm sech}^2\left(\frac{x}{\xi }\right),
\end{align}
for the continuum state $|E|>\Delta_0\sin\theta$.

The $\theta$-angle-resolved mass current at the edge for $\theta=\pi/2$ is shown in Fig.~\ref{energy}(b).
The Majorana contribution to the current is linear in $E$.
The edge mass current in the continuum state decreases away from $|\Delta_0\sin\theta|$.
The asymptotic behavior of the edge current (11) at $x=0$
is estimated as
\begin{align}
j_y(x=0,\theta, E)\approx -{mv_FN_0\over 16}\sin\theta\left(\frac{\Delta_0\sin\theta}{E}\right)^3,
\end{align}
for $E\ll -\Delta_0\sin\theta$, implying that the contribution
of the edge current decreases $\sim|E|^{-3}$.
This power law behavior of the decrease means that it is 
neither confined near $E_F$, nor deep up to the band bottom,
rather  it is marginal.

Since QPs fill the energy state up to the Fermi energy at the zero temperature,
the mass current along the edge from the Majorana bound state is obtained analytically as
\begin{align}
j_y^{\rm MJ}(x)&=\left\langle\int_{-\Delta_0\sin\theta }^{0}dEj_y(x,\bi{k},E)\right\rangle_{\bi{k}}\nn\\
&=-\frac{mv_FN_0\Delta_0}{6}\ {\rm sech}^2\left(\frac{x}{\xi }\right),
\label{jMJ}
\end{align}
and that from the continuum state is 
\begin{align}
j_y^{\rm cont}(x)&=\left\langle\int_{-\infty }^{-\Delta_0\sin\theta }dEj_y(x,\bi{k},E)\right\rangle_{\bi{k}}\nn\\
&=\frac{mv_FN_0\Delta_0}{12}\ {\rm sech}^2\left(\frac{x}{\xi }\right),
\end{align}
where a similar result to Eq.~\eqref{jMJ} is obtained in connection with the chiral superconductor Sr$_2$RuO$_4$~\cite{furusaki:2001}.
Thus those currents flow oppositely.
In the disk with $R\gg\xi$, the angular momentum by the edge mass current from each state is
calculated as 
\begin{align}
L_z^{\rm MJ}=N\hbar,\ L_z^{\rm cont}=-\frac{N\hbar }{2},
\end{align}
The factor 2 difference in those contributions may be related
to our finding that the DOS of the Majorana QPs is $N_0/2$.
Finally, the total angular momentum becomes simply to
\begin{align}
L_z=L_z^{\rm MJ}+L_z^{\rm cont}=\frac{N\hbar }{2},
\end{align}
which corresponds to the intrinsic angular momentum for $\gamma=0$ mentioned above.
The total angular momentum due to the edge mass current coincides with that in Ref.~11.
Here we emphasize that we decompose it analytically into two contributions:
the Majorana bound state and continuum state.
Interestingly, a half of the angular momentum from the Majorana 
bound state is canceled partly by that from the continuum state,
resulting in the expected angular momentum.
Note that the total number of $^3$He atoms $N$ emerges from the normal DOS 
$N_0=(3/mv_F^2)(N/V)$, where $V$ is the volume of the disk.

\begin{figure}
\begin{center}
\includegraphics[width=6cm]{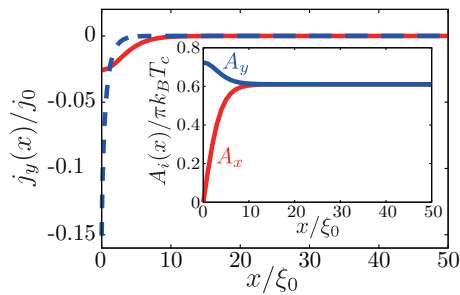}
\end{center}
\caption{\label{profile}(Color online)
Profiles of mass current along the edge at $T=0.5T_c$ under self-consistent pair potential (solid line)
and uniform pair potential (dashed line).
The units are $\xi_0=\hbar v_F/2\pi k_BT_c$ and $j_0=mv_FN_0\pi k_BT_c$.
Inset: A profile of the self-consistent pair potential.}
\end{figure}

Next, we solve Eilenberger equation~\eqref{eilenberger} and gap equation~\eqref{gap} self-consistently
and compare to the solution under the uniform pair potential.
The uniform pair potential means $\Delta(x,\bi{k})=\Delta_0(k_x+ik_y)$,
where amplitude of $\Delta_0$ is determined by the gap equation.
In Fig.~\ref{profile}, the profiles of the edge mass current at $T=0.5T_c$ under self-consistent pair potential (solid line)
and uniform pair potential (dashed line) calculated by Eq.~\eqref{current} are shown.
The profile of the edge mass current under the self-consistent pair potential is 
varied gradually over the region within $\sim 10\xi_0$
owing to the variation of the pair potential (the inset of Fig.~\ref{profile}).
In contrast the edge mass current under the uniform pair potential is localized near the edge.
A profile of the self-consistent pair potential is also shown in the inset of Fig.~\ref{profile}.
Because of the specular boundary condition, the $k_x$-component becomes zero at the edge.
The $k_y$-component is enhanced by compensating for the loss of the $k_x$-component on the edge,
where the polar state is realized.
In other words, the inverse chiral state $-k_x+ik_y$ is mixed with the chiral state $k_x+ik_y$ near the edge.
This inverse chiral state must have the $4\pi$ phase winding 
to minimize the free energy in the axisymmetric disk~\cite{tsutsumi:2008}.

\begin{figure}
\begin{center}
\includegraphics[width=6.5cm]{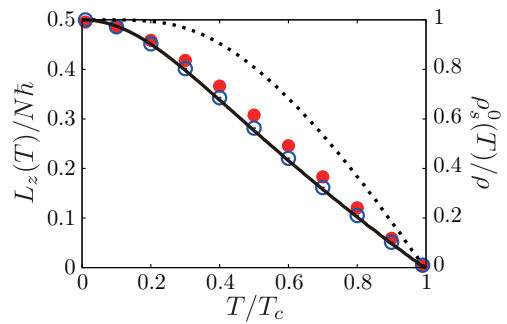}
\end{center}
\caption{\label{temperature}(Color online)
Left axis: Temperature dependence of angular momentum under self-consistent pair potential (solid circles)
and uniform pair potential (open circles).
Right axis: Temperature dependence of the component of the superfluid density tensor $\rho_{s\parallel }^0$ (solid line) 
and $\rho_{s\perp }^0$ (dotted line).}
\end{figure}

The temperature dependence of the angular momentum $L_z(T)$ by the edge mass current 
under the above self-consistent and the uniform pair potential is shown in Fig.~\ref{temperature} by circles.
The component of the superfluid density tensor 
parallel (perpendicular) to the direction of the point nodes $\rho_{s\parallel }^0$ ($\rho_{s\perp }^0$)~\cite{cross:1975} is also depicted by two lines.
In both pair potential cases, $L_z$ tends to $N\hbar/2$ toward the low temperature limit 
and zero at the transition temperature.
In the uniform case, $L_z(T)$ (open circles) has the same temperature dependence as $\rho_{s\parallel }^0(T)$ (solid line)~\cite{note2},
because thermal excitations from the point nodes reduce the superfluid density.
However, $L_z(T)$ in the self-consistent case (solid circles) is larger than
that in the uniform case (open circles) in intermediate temperatures.
The reason for the increment of $L_z(T)$ is that the additional 
inverse chiral state with the $4\pi$ phase winding yields the extra current.
The different $T$-dependences between $L_z(T)$ and $\rho_{s\parallel }^0(T)$
should be observed experimentally in wide intermediate $T$-region.

According to the above analytic solution~\eqref{Green}, 
we can derive the low temperature behavior of the total angular momentum as
\begin{align}
L_z(T)\approx {N\hbar\over 2}\left[1-\beta\left({\pi k_BT\over \Delta_0}\right)^2+\mathrm{O}\left({\pi k_BT\over \Delta_0}\right)^4\right],
\end{align}
with $\beta=1$. 
Note that $\beta=2/3$ in the absence of point nodes, 
namely in the two-dimensional Fermi surface model.
The $T^2$-dependence comes from the Majorana QPs
because $j_y(x,\theta,E)$ in the Majorana bound state has linear energy dependence (Fig.~\ref{energy}(b)).
Thus, the observation of the $T^2$-behavior could establish the existence of the Majorana QPs.

We point out another possibility to experimentally observe the Majorana QPs contribution $L^{\rm MJ}$ and the continuum contribution $L^{\rm cont}$ separately by controlling the surface roughness.
The surface roughness can be controlled experimentally by coating the surface with $^4$He atoms~\cite{wada:2008}.
The angular momenta $L^{\rm MJ}$ and $L^{\rm cont}$ can be suppressed by the different rates as a function of the surface roughness. 
Here we notice in the case of the superfluid $^3$He B-phase 
that the surface Majorana bound state remains at a partially specular surface~\cite{murakawa:2009,murakawa:2011}.
The concrete calculation to estimate quantitatively this change is a future problem.

Recently a paper by Sauls~\cite{sauls:2011} has appeared, in which the author discusses the same topics as our own.
However, he uses the two-dimensional Fermi surface model under the uniform pair potential. 
We reach the same conclusions regarding the angular momentum at zero temperature with the specular boundary.

In summary, we have found an analytic solution~\eqref{Green} of Eilenberger equation 
for the superfluid $^3$He A-phase at the boundary.
This analytic solution is useful enough to elucidate the elusive Majorana QPs
in a realistic situation accessible experimentally.
Namely, the analytic expressions for the DOS, the mass current and the 
associated angular momentum of the disk, that make clear the roles of the Majorana QPs.
We have succeeded in separating $L=N\hbar/2$ into $L^{\rm MJ}=N\hbar$
and $L^{\rm cont}=-N\hbar/2$ and also in proving the DOS of the Majorana QPs
is exactly the half of the normal DOS $N_0/2$.
The self-consistent numerical computation is consistent with those conclusions 
and provides further information on the Majorana QPs that facilitates future experiment 
for detecting this exotic particle.

Finally, we mention that the relation between the intrinsic angular momentum and the edge mass current.
The mass current by a Cooper pair is cancelled by overlapping Cooper pairs in the bulk.
The total current is zero by the particle-hole symmetry~\cite{cross:1975}.
At the edge, however, since the cancellation does not occur completely,
the angular momentum $L=N\hbar/2$ emerges via the edge mass current.
The superfluid $^3$He A-phase in disk geometry has the net angular momentum $L=N\hbar/2$.
In three-dimensional thick slabs, the net angular momentum reduces by the canting of $l$-vector at the edge~\cite{tsutsumi:2011b}.

We thank M.~Ichioka, T.~Mizushima, K.~Nagai, S.~Higashitani, and Y.~Nagato for helpful discussions.
Y.T.~acknowledges the support of the Research Fellowships of the Japan Society for the Promotion of Science for Young Scientists.


\begin{thebibliography}{25}%
\makeatletter
\providecommand \@ifxundefined [1]{%
 \@ifx{#1\undefined}
}%
\providecommand \@ifnum [1]{%
 \ifnum #1\expandafter \@firstoftwo
 \else \expandafter \@secondoftwo
 \fi
}%
\providecommand \@ifx [1]{%
 \ifx #1\expandafter \@firstoftwo
 \else \expandafter \@secondoftwo
 \fi
}%
\providecommand \natexlab [1]{#1}%
\providecommand \enquote  [1]{``#1''}%
\providecommand \bibnamefont  [1]{#1}%
\providecommand \bibfnamefont [1]{#1}%
\providecommand \citenamefont [1]{#1}%
\providecommand \href@noop [0]{\@secondoftwo}%
\providecommand \href [0]{\begingroup \@sanitize@url \@href}%
\providecommand \@href[1]{\@@startlink{#1}\@@href}%
\providecommand \@@href[1]{\endgroup#1\@@endlink}%
\providecommand \@sanitize@url [0]{\catcode `\\12\catcode `\$12\catcode
  `\&12\catcode `\#12\catcode `\^12\catcode `\_12\catcode `\%12\relax}%
\providecommand \@@startlink[1]{}%
\providecommand \@@endlink[0]{}%
\providecommand \url  [0]{\begingroup\@sanitize@url \@url }%
\providecommand \@url [1]{\endgroup\@href {#1}{\urlprefix }}%
\providecommand \urlprefix  [0]{URL }%
\providecommand \Eprint [0]{\href }%
\providecommand \doibase [0]{http://dx.doi.org/}%
\providecommand \selectlanguage [0]{\@gobble}%
\providecommand \bibinfo  [0]{\@secondoftwo}%
\providecommand \bibfield  [0]{\@secondoftwo}%
\providecommand \translation [1]{[#1]}%
\providecommand \BibitemOpen [0]{}%
\providecommand \bibitemStop [0]{}%
\providecommand \bibitemNoStop [0]{.\EOS\space}%
\providecommand \EOS [0]{\spacefactor3000\relax}%
\providecommand \BibitemShut  [1]{\csname bibitem#1\endcsname}%
\let\auto@bib@innerbib\@empty
\bibitem [{\citenamefont {Vollhardt}\ and\ \citenamefont
  {W{\"o}lfle}(1990)}]{vollhardt:book}%
  \BibitemOpen
  \bibfield  {author} {\bibinfo {author} {\bibfnamefont {D.}~\bibnamefont
  {Vollhardt}}\ and\ \bibinfo {author} {\bibfnamefont {P.}~\bibnamefont
  {W{\"o}lfle}},\ }\href@noop {} {\emph {\bibinfo {title} {The Superfluid Phase
  of Helium 3}}}\ (\bibinfo  {publisher} {Taylor and Francis},\ \bibinfo
  {address} {London},\ \bibinfo {year} {1990})\BibitemShut {NoStop}%
\bibitem [{\citenamefont {Hasan}\ and\ \citenamefont
  {Kane}(2010)}]{hasan:2010}%
  \BibitemOpen
  \bibfield  {author} {\bibinfo {author} {\bibfnamefont {M.~Z.}\ \bibnamefont
  {Hasan}}\ and\ \bibinfo {author} {\bibfnamefont {C.~L.}\ \bibnamefont
  {Kane}},\ }\href {\doibase 10.1103/RevModPhys.82.3045} {\bibfield  {journal}
  {\bibinfo  {journal} {Rev. Mod. Phys.}\ }\textbf {\bibinfo {volume} {82}},\
  \bibinfo {pages} {3045} (\bibinfo {year} {2010})}\BibitemShut {NoStop}%
\bibitem [{\citenamefont {Read}\ and\ \citenamefont {Green}(2000)}]{read:2000}%
  \BibitemOpen
  \bibfield  {author} {\bibinfo {author} {\bibfnamefont {N.}~\bibnamefont
  {Read}}\ and\ \bibinfo {author} {\bibfnamefont {D.}~\bibnamefont {Green}},\
  }\href@noop {} {\bibfield  {journal} {\bibinfo  {journal} {Phys. Rev. B}\
  }\textbf {\bibinfo {volume} {61}},\ \bibinfo {pages} {10267} (\bibinfo {year}
  {2000})}\BibitemShut {NoStop}%
\bibitem [{\citenamefont {Tsutsumi}\ \emph {et~al.}(2011)\citenamefont
  {Tsutsumi}, \citenamefont {Ichioka},\ and\ \citenamefont
  {Machida}}]{tsutsumi:2011b}%
  \BibitemOpen
  \bibfield  {author} {\bibinfo {author} {\bibfnamefont {Y.}~\bibnamefont
  {Tsutsumi}}, \bibinfo {author} {\bibfnamefont {M.}~\bibnamefont {Ichioka}}, \
  and\ \bibinfo {author} {\bibfnamefont {K.}~\bibnamefont {Machida}},\ }\href
  {\doibase 10.1103/PhysRevB.83.094510} {\bibfield  {journal} {\bibinfo
  {journal} {Phys. Rev. B}\ }\textbf {\bibinfo {volume} {83}},\ \bibinfo
  {pages} {094510} (\bibinfo {year} {2011})}\BibitemShut {NoStop}%
\bibitem [{\citenamefont {Tsutsumi}\ \emph {et~al.}(2010)\citenamefont
  {Tsutsumi}, \citenamefont {Mizushima}, \citenamefont {Ichioka},\ and\
  \citenamefont {Machida}}]{tsutsumi:2010b}%
  \BibitemOpen
  \bibfield  {author} {\bibinfo {author} {\bibfnamefont {Y.}~\bibnamefont
  {Tsutsumi}}, \bibinfo {author} {\bibfnamefont {T.}~\bibnamefont {Mizushima}},
  \bibinfo {author} {\bibfnamefont {M.}~\bibnamefont {Ichioka}}, \ and\
  \bibinfo {author} {\bibfnamefont {K.}~\bibnamefont {Machida}},\ }\href@noop
  {} {\bibfield  {journal} {\bibinfo  {journal} {J. Phys. Soc. Jpn.}\ }\textbf
  {\bibinfo {volume} {79}},\ \bibinfo {pages} {113601} (\bibinfo {year}
  {2010})}\BibitemShut {NoStop}%
\bibitem [{\citenamefont {Schnyder}\ \emph {et~al.}(2008)\citenamefont
  {Schnyder}, \citenamefont {Ryu}, \citenamefont {Furusaki},\ and\
  \citenamefont {Ludwig}}]{schnyder:2008}%
  \BibitemOpen
  \bibfield  {author} {\bibinfo {author} {\bibfnamefont {A.~P.}\ \bibnamefont
  {Schnyder}}, \bibinfo {author} {\bibfnamefont {S.}~\bibnamefont {Ryu}},
  \bibinfo {author} {\bibfnamefont {A.}~\bibnamefont {Furusaki}}, \ and\
  \bibinfo {author} {\bibfnamefont {A.~W.~W.}\ \bibnamefont {Ludwig}},\ }\href
  {\doibase 10.1103/PhysRevB.78.195125} {\bibfield  {journal} {\bibinfo
  {journal} {Phys. Rev. B}\ }\textbf {\bibinfo {volume} {78}},\ \bibinfo
  {pages} {195125} (\bibinfo {year} {2008})}\BibitemShut {NoStop}%
\bibitem [{\citenamefont {Chung}\ and\ \citenamefont
  {Zhang}(2009)}]{chung:2009}%
  \BibitemOpen
  \bibfield  {author} {\bibinfo {author} {\bibfnamefont {S.~B.}\ \bibnamefont
  {Chung}}\ and\ \bibinfo {author} {\bibfnamefont {S.-C.}\ \bibnamefont
  {Zhang}},\ }\href@noop {} {\bibfield  {journal} {\bibinfo  {journal} {Phys.
  Rev. Lett.}\ }\textbf {\bibinfo {volume} {103}},\ \bibinfo {pages} {235301}
  (\bibinfo {year} {2009})}\BibitemShut {NoStop}%
\bibitem [{\citenamefont {Nagato}\ \emph {et~al.}(2009)\citenamefont {Nagato},
  \citenamefont {Higashitani},\ and\ \citenamefont {Nagai}}]{nagato:2009}%
  \BibitemOpen
  \bibfield  {author} {\bibinfo {author} {\bibfnamefont {Y.}~\bibnamefont
  {Nagato}}, \bibinfo {author} {\bibfnamefont {S.}~\bibnamefont {Higashitani}},
  \ and\ \bibinfo {author} {\bibfnamefont {K.}~\bibnamefont {Nagai}},\
  }\href@noop {} {\bibfield  {journal} {\bibinfo  {journal} {J. Phys. Soc.
  Jpn.}\ }\textbf {\bibinfo {volume} {78}},\ \bibinfo {pages} {123603}
  (\bibinfo {year} {2009})}\BibitemShut {NoStop}%
\bibitem [{\citenamefont {Murakawa}\ \emph {et~al.}(2009)\citenamefont
  {Murakawa}, \citenamefont {Tamura}, \citenamefont {Wada}, \citenamefont
  {Wasai}, \citenamefont {Saitoh}, \citenamefont {Aoki}, \citenamefont
  {Nomura}, \citenamefont {Okuda}, \citenamefont {Nagato}, \citenamefont
  {Yamamoto}, \citenamefont {Higashitani},\ and\ \citenamefont
  {Nagai}}]{murakawa:2009}%
  \BibitemOpen
  \bibfield  {author} {\bibinfo {author} {\bibfnamefont {S.}~\bibnamefont
  {Murakawa}}, \bibinfo {author} {\bibfnamefont {Y.}~\bibnamefont {Tamura}},
  \bibinfo {author} {\bibfnamefont {Y.}~\bibnamefont {Wada}}, \bibinfo {author}
  {\bibfnamefont {M.}~\bibnamefont {Wasai}}, \bibinfo {author} {\bibfnamefont
  {M.}~\bibnamefont {Saitoh}}, \bibinfo {author} {\bibfnamefont
  {Y.}~\bibnamefont {Aoki}}, \bibinfo {author} {\bibfnamefont {R.}~\bibnamefont
  {Nomura}}, \bibinfo {author} {\bibfnamefont {Y.}~\bibnamefont {Okuda}},
  \bibinfo {author} {\bibfnamefont {Y.}~\bibnamefont {Nagato}}, \bibinfo
  {author} {\bibfnamefont {M.}~\bibnamefont {Yamamoto}}, \bibinfo {author}
  {\bibfnamefont {S.}~\bibnamefont {Higashitani}}, \ and\ \bibinfo {author}
  {\bibfnamefont {K.}~\bibnamefont {Nagai}},\ }\href {\doibase
  10.1103/PhysRevLett.103.155301} {\bibfield  {journal} {\bibinfo  {journal}
  {Phys. Rev. Lett.}\ }\textbf {\bibinfo {volume} {103}},\ \bibinfo {pages}
  {155301} (\bibinfo {year} {2009})}\BibitemShut {NoStop}%
\bibitem [{\citenamefont {Murakawa}\ \emph {et~al.}(2011)\citenamefont
  {Murakawa}, \citenamefont {Wada}, \citenamefont {Tamura}, \citenamefont
  {Wasai}, \citenamefont {Saitoh}, \citenamefont {Aoki}, \citenamefont
  {Nomura}, \citenamefont {Okuda}, \citenamefont {Nagato}, \citenamefont
  {Yamamoto}, \citenamefont {Higashitani},\ and\ \citenamefont
  {Nagai}}]{murakawa:2011}%
  \BibitemOpen
  \bibfield  {author} {\bibinfo {author} {\bibfnamefont {S.}~\bibnamefont
  {Murakawa}}, \bibinfo {author} {\bibfnamefont {Y.}~\bibnamefont {Wada}},
  \bibinfo {author} {\bibfnamefont {Y.}~\bibnamefont {Tamura}}, \bibinfo
  {author} {\bibfnamefont {M.}~\bibnamefont {Wasai}}, \bibinfo {author}
  {\bibfnamefont {M.}~\bibnamefont {Saitoh}}, \bibinfo {author} {\bibfnamefont
  {Y.}~\bibnamefont {Aoki}}, \bibinfo {author} {\bibfnamefont {R.}~\bibnamefont
  {Nomura}}, \bibinfo {author} {\bibfnamefont {Y.}~\bibnamefont {Okuda}},
  \bibinfo {author} {\bibfnamefont {Y.}~\bibnamefont {Nagato}}, \bibinfo
  {author} {\bibfnamefont {M.}~\bibnamefont {Yamamoto}}, \bibinfo {author}
  {\bibfnamefont {S.}~\bibnamefont {Higashitani}}, \ and\ \bibinfo {author}
  {\bibfnamefont {K.}~\bibnamefont {Nagai}},\ }\href {\doibase
  10.1143/JPSJ.80.013602} {\bibfield  {journal} {\bibinfo  {journal} {J. Phys.
  Soc. Jpn.}\ }\textbf {\bibinfo {volume} {80}},\ \bibinfo {pages} {013602}
  (\bibinfo {year} {2011})}\BibitemShut {NoStop}%
\bibitem [{\citenamefont {Stone}\ and\ \citenamefont {Roy}(2004)}]{stone:2004}%
  \BibitemOpen
  \bibfield  {author} {\bibinfo {author} {\bibfnamefont {M.}~\bibnamefont
  {Stone}}\ and\ \bibinfo {author} {\bibfnamefont {R.}~\bibnamefont {Roy}},\
  }\href@noop {} {\bibfield  {journal} {\bibinfo  {journal} {Phys. Rev. B}\
  }\textbf {\bibinfo {volume} {69}},\ \bibinfo {pages} {184511} (\bibinfo
  {year} {2004})}\BibitemShut {NoStop}%
\bibitem [{\citenamefont {Ishikawa}(1977)}]{ishikawa:1977}%
  \BibitemOpen
  \bibfield  {author} {\bibinfo {author} {\bibfnamefont {M.}~\bibnamefont
  {Ishikawa}},\ }\href@noop {} {\bibfield  {journal} {\bibinfo  {journal}
  {Prog. Theor. Phys.}\ }\textbf {\bibinfo {volume} {57}},\ \bibinfo {pages}
  {1836} (\bibinfo {year} {1977})}\BibitemShut {NoStop}%
\bibitem [{\citenamefont {Ishikawa}\ \emph {et~al.}(1980)\citenamefont
  {Ishikawa}, \citenamefont {Miyake},\ and\ \citenamefont
  {Usui}}]{ishikawa:1980}%
  \BibitemOpen
  \bibfield  {author} {\bibinfo {author} {\bibfnamefont {M.}~\bibnamefont
  {Ishikawa}}, \bibinfo {author} {\bibfnamefont {K.}~\bibnamefont {Miyake}}, \
  and\ \bibinfo {author} {\bibfnamefont {T.}~\bibnamefont {Usui}},\ }\href@noop
  {} {\bibfield  {journal} {\bibinfo  {journal} {Prog. Theor. Phys.}\ }\textbf
  {\bibinfo {volume} {63}},\ \bibinfo {pages} {1083} (\bibinfo {year}
  {1980})}\BibitemShut {NoStop}%
\bibitem [{\citenamefont {Mermin}\ and\ \citenamefont
  {Muzikar}(1980)}]{mermin:1980}%
  \BibitemOpen
  \bibfield  {author} {\bibinfo {author} {\bibfnamefont {N.~D.}\ \bibnamefont
  {Mermin}}\ and\ \bibinfo {author} {\bibfnamefont {P.}~\bibnamefont
  {Muzikar}},\ }\href {\doibase 10.1103/PhysRevB.21.980} {\bibfield  {journal}
  {\bibinfo  {journal} {Phys. Rev. B}\ }\textbf {\bibinfo {volume} {21}},\
  \bibinfo {pages} {980} (\bibinfo {year} {1980})}\BibitemShut {NoStop}%
\bibitem [{\citenamefont {Volovik}(1995)}]{volovik:1995}%
  \BibitemOpen
  \bibfield  {author} {\bibinfo {author} {\bibfnamefont {G.~E.}\ \bibnamefont
  {Volovik}},\ }\href@noop {} {\bibfield  {journal} {\bibinfo  {journal} {JETP
  Lett.}\ }\textbf {\bibinfo {volume} {61}},\ \bibinfo {pages} {958} (\bibinfo
  {year} {1995})}\BibitemShut {NoStop}%
\bibitem [{\citenamefont {Kita}(1998)}]{kita:1998}%
  \BibitemOpen
  \bibfield  {author} {\bibinfo {author} {\bibfnamefont {T.}~\bibnamefont
  {Kita}},\ }\href@noop {} {\bibfield  {journal} {\bibinfo  {journal} {J. Phys.
  Soc. Jpn.}\ }\textbf {\bibinfo {volume} {67}},\ \bibinfo {pages} {216}
  (\bibinfo {year} {1998})}\BibitemShut {NoStop}%
\bibitem [{\citenamefont {Anderson}\ and\ \citenamefont
  {Morel}(1961)}]{anderson:1961}%
  \BibitemOpen
  \bibfield  {author} {\bibinfo {author} {\bibfnamefont {P.~W.}\ \bibnamefont
  {Anderson}}\ and\ \bibinfo {author} {\bibfnamefont {P.}~\bibnamefont
  {Morel}},\ }\href {\doibase 10.1103/PhysRev.123.1911} {\bibfield  {journal}
  {\bibinfo  {journal} {Phys. Rev.}\ }\textbf {\bibinfo {volume} {123}},\
  \bibinfo {pages} {1911} (\bibinfo {year} {1961})}\BibitemShut {NoStop}%
\bibitem [{\citenamefont {Leggett}(1975)}]{leggett:1975}%
  \BibitemOpen
  \bibfield  {author} {\bibinfo {author} {\bibfnamefont {A.~J.}\ \bibnamefont
  {Leggett}},\ }\href@noop {} {\bibfield  {journal} {\bibinfo  {journal} {Rev.
  Mod. Phys.}\ }\textbf {\bibinfo {volume} {47}},\ \bibinfo {pages} {331}
  (\bibinfo {year} {1975})}\BibitemShut {NoStop}%
\bibitem [{\citenamefont {Volovik}(1975)}]{volovik:1975a}%
  \BibitemOpen
  \bibfield  {author} {\bibinfo {author} {\bibfnamefont {G.~E.}\ \bibnamefont
  {Volovik}},\ }\href@noop {} {\bibfield  {journal} {\bibinfo  {journal} {JETP
  Lett.}\ }\textbf {\bibinfo {volume} {22}},\ \bibinfo {pages} {108} (\bibinfo
  {year} {1975})}\BibitemShut {NoStop}%
\bibitem [{\citenamefont {Cross}(1975)}]{cross:1975}%
  \BibitemOpen
  \bibfield  {author} {\bibinfo {author} {\bibfnamefont {M.}~\bibnamefont
  {Cross}},\ }\href@noop {} {\bibfield  {journal} {\bibinfo  {journal} {J. Low
  Temp. Phys.}\ }\textbf {\bibinfo {volume} {21}},\ \bibinfo {pages} {525}
  (\bibinfo {year} {1975})}\BibitemShut {NoStop}%
\bibitem [{\citenamefont {Bennett}\ \emph {et~al.}(2010)\citenamefont
  {Bennett}, \citenamefont {Levitin}, \citenamefont {Casey}, \citenamefont
  {Cowan}, \citenamefont {Parpia},\ and\ \citenamefont
  {Saunders}}]{bennett:2010}%
  \BibitemOpen
  \bibfield  {author} {\bibinfo {author} {\bibfnamefont {R.~G.}\ \bibnamefont
  {Bennett}}, \bibinfo {author} {\bibfnamefont {L.~V.}\ \bibnamefont
  {Levitin}}, \bibinfo {author} {\bibfnamefont {A.}~\bibnamefont {Casey}},
  \bibinfo {author} {\bibfnamefont {B.}~\bibnamefont {Cowan}}, \bibinfo
  {author} {\bibfnamefont {J.}~\bibnamefont {Parpia}}, \ and\ \bibinfo {author}
  {\bibfnamefont {J.}~\bibnamefont {Saunders}},\ }\href@noop {} {\bibfield
  {journal} {\bibinfo  {journal} {J. Low Temp. Phys.}\ }\textbf {\bibinfo
  {volume} {158}},\ \bibinfo {pages} {163} (\bibinfo {year}
  {2010})}\BibitemShut {NoStop}%
\bibitem [{\citenamefont {Eilenberger}(1968)}]{eilenberger:1968}%
  \BibitemOpen
  \bibfield  {author} {\bibinfo {author} {\bibfnamefont {G.}~\bibnamefont
  {Eilenberger}},\ }\href@noop {} {\bibfield  {journal} {\bibinfo  {journal}
  {Z. Phys.}\ }\textbf {\bibinfo {volume} {214}},\ \bibinfo {pages} {195}
  (\bibinfo {year} {1968})}\BibitemShut {NoStop}%
\bibitem [{\citenamefont {Nagato}\ \emph {et~al.}(1993)\citenamefont {Nagato},
  \citenamefont {Nagai},\ and\ \citenamefont {Hara}}]{nagato:1993}%
  \BibitemOpen
  \bibfield  {author} {\bibinfo {author} {\bibfnamefont {Y.}~\bibnamefont
  {Nagato}}, \bibinfo {author} {\bibfnamefont {K.}~\bibnamefont {Nagai}}, \
  and\ \bibinfo {author} {\bibfnamefont {J.}~\bibnamefont {Hara}},\ }\href@noop
  {} {\bibfield  {journal} {\bibinfo  {journal} {J. Low Temp. Phys.}\ }\textbf
  {\bibinfo {volume} {93}},\ \bibinfo {pages} {33} (\bibinfo {year}
  {1993})}\BibitemShut {NoStop}%
\bibitem [{\citenamefont {Schopohl}\ and\ \citenamefont
  {Maki}(1995)}]{schopohl:1995}%
  \BibitemOpen
  \bibfield  {author} {\bibinfo {author} {\bibfnamefont {N.}~\bibnamefont
  {Schopohl}}\ and\ \bibinfo {author} {\bibfnamefont {K.}~\bibnamefont
  {Maki}},\ }\href {\doibase 10.1103/PhysRevB.52.490} {\bibfield  {journal}
  {\bibinfo  {journal} {Phys. Rev. B}\ }\textbf {\bibinfo {volume} {52}},\
  \bibinfo {pages} {490} (\bibinfo {year} {1995})}\BibitemShut {NoStop}%
\bibitem [{\citenamefont {Schopohl}()}]{schopohl:cond}%
  \BibitemOpen
  \bibfield  {author} {\bibinfo {author} {\bibfnamefont {N.}~\bibnamefont
  {Schopohl}},\ }\href@noop {} {\bibinfo  {journal} {arXiv:cond-mat/9804064}\
  }\BibitemShut {NoStop}%
\bibitem{note1}
  We note that this analytical solution is shown to be self-consistent one in a certain limit,
  namely the energy cutoff $\omega_c$ becomes infinite.
\bibitem [{\citenamefont {Furusaki}\ \emph {et~al.}(2001)\citenamefont
  {Furusaki}, \citenamefont {Matsumoto},\ and\ \citenamefont
  {Sigrist}}]{furusaki:2001}%
  \BibitemOpen
\bibfield  {journal} {  }\bibfield  {author} {\bibinfo {author} {\bibfnamefont
  {A.}~\bibnamefont {Furusaki}}, \bibinfo {author} {\bibfnamefont
  {M.}~\bibnamefont {Matsumoto}}, \ and\ \bibinfo {author} {\bibfnamefont
  {M.}~\bibnamefont {Sigrist}},\ }\href {\doibase 10.1103/PhysRevB.64.054514}
  {\bibfield  {journal} {\bibinfo  {journal} {Phys. Rev. B}\ }\textbf {\bibinfo
  {volume} {64}},\ \bibinfo {pages} {054514} (\bibinfo {year}
  {2001})}\BibitemShut {NoStop}%
\bibitem [{\citenamefont {Tsutsumi}\ \emph {et~al.}(2008)\citenamefont
  {Tsutsumi}, \citenamefont {Kawakami}, \citenamefont {Mizushima},
  \citenamefont {Ichioka},\ and\ \citenamefont {Machida}}]{tsutsumi:2008}%
  \BibitemOpen
  \bibfield  {author} {\bibinfo {author} {\bibfnamefont {Y.}~\bibnamefont
  {Tsutsumi}}, \bibinfo {author} {\bibfnamefont {T.}~\bibnamefont {Kawakami}},
  \bibinfo {author} {\bibfnamefont {T.}~\bibnamefont {Mizushima}}, \bibinfo
  {author} {\bibfnamefont {M.}~\bibnamefont {Ichioka}}, \ and\ \bibinfo
  {author} {\bibfnamefont {K.}~\bibnamefont {Machida}},\ }\href@noop {}
  {\bibfield  {journal} {\bibinfo  {journal} {Phys. Rev. Lett.}\ }\textbf
  {\bibinfo {volume} {101}},\ \bibinfo {pages} {135302} (\bibinfo {year}
  {2008})}\BibitemShut {NoStop}%
\bibitem{note2}
  This fact was pointed in Ref.~16.
\bibitem [{\citenamefont {Wada}\ \emph {et~al.}(2008)\citenamefont {Wada},
  \citenamefont {Murakawa}, \citenamefont {Tamura}, \citenamefont {Saitoh},
  \citenamefont {Aoki}, \citenamefont {Nomura},\ and\ \citenamefont
  {Okuda}}]{wada:2008}%
  \BibitemOpen
  \bibfield  {author} {\bibinfo {author} {\bibfnamefont {Y.}~\bibnamefont
  {Wada}}, \bibinfo {author} {\bibfnamefont {S.}~\bibnamefont {Murakawa}},
  \bibinfo {author} {\bibfnamefont {Y.}~\bibnamefont {Tamura}}, \bibinfo
  {author} {\bibfnamefont {M.}~\bibnamefont {Saitoh}}, \bibinfo {author}
  {\bibfnamefont {Y.}~\bibnamefont {Aoki}}, \bibinfo {author} {\bibfnamefont
  {R.}~\bibnamefont {Nomura}}, \ and\ \bibinfo {author} {\bibfnamefont
  {Y.}~\bibnamefont {Okuda}},\ }\href {\doibase 10.1103/PhysRevB.78.214516}
  {\bibfield  {journal} {\bibinfo  {journal} {Phys. Rev. B}\ }\textbf {\bibinfo
  {volume} {78}},\ \bibinfo {pages} {214516} (\bibinfo {year}
  {2008})}\BibitemShut {NoStop}%
\bibitem [{\citenamefont {Sauls}(2011)}]{sauls:2011}%
  \BibitemOpen
  \bibfield  {author} {\bibinfo {author} {\bibfnamefont {J.~A.}\ \bibnamefont
  {Sauls}},\ }\href {\doibase 10.1103/PhysRevB.84.214509} {\bibfield  {journal}
  {\bibinfo  {journal} {Phys. Rev. B}\ }\textbf {\bibinfo {volume} {84}},\
  \bibinfo {pages} {214509} (\bibinfo {year} {2011})}\BibitemShut {NoStop}%
\end{thebibliography}

%

\end{document}